\documentclass[%
 reprint,
 showpacs,preprintnumbers,
 nofootinbib,
 amsmath,amssymb,
 aps,
]{revtex4-1}

\usepackage{graphicx}
\usepackage{dcolumn}
\usepackage{bm}

\usepackage{amsmath}
\usepackage{ulem}
\usepackage[usenames]{color}
\usepackage{comment}

\begin{document}

\title{Hybrid stars from the NJL model with a tensor interaction}

\author{Hiroaki Matsuoka}%
\email{b16d6a01@s.kochi-u.ac.jp}
\affiliation{Graduate School of Integrated Arts and Science, Kochi University, Kochi 780-8520, Japan}

\author{Yasuhiko Tsue}
\email{tsue@kochi-u.ac.jp}
\affiliation{Department of Mathematics and Physics, Kochi University, Kochi 780-8520, Japan}

\author{Jo\~{a}o da Provid\^{e}ncia}%
\author{Constan\c{c}a Provid\^{e}ncia}
\affiliation{CFisUC, Departamento de F\'{i}sica, Universidade de Coimbra, 3004-516 Coimbra, Portugal}

\author{Masatoshi Yamamura}
\affiliation{Department of Pure and Applied Physics, Faculty of Engineering Science, Kansai University,
Suita 564-8680, Japan}

\date{\today}

\begin{abstract}
In order to obtain the equation of state and construct hybrid stars, we calculate the thermodynamic potential 
in the two-flavor Nambu--Jona-Lasinio model with tensor-type four-point interaction between quarks.
In addition, we impose the $\beta$ equilibrium and charge neutrality conditions on the system.
We show that the tensor condensate appears at large chemical potential, however, it is difficult to hold hybrid stars with two-solar mass by using the equation of state with the tensor interaction.
Although we cannot obtain the stars with two-solar mass because of the absence of the repulsive interaction, the estimated magnetic moment density is very large.
Therefore, we expect that the tensor interaction describes the magnetic fields of compact stars.
\end{abstract}



\maketitle


\section{Introduction}

One of recent interests is to clarify 
the phase structure of the world governed by the quantum chromodynamics (QCD).
It is well known that at low temperature and small chemical potential the hadronic phase is realized.
In this phase, because of the color confinement, we cannot remove a single quark from hadrons.
At high temperature and small chemical potential, the quark-gluon plasma phase may be realized.
On the other hand, it is very difficult to investigate the phase structure at low temperature and large chemical potential.
Many researchers, however, consider that the color superconducting phase may be realized under certain conditions \cite{Alford:2001}.

In order to exhibit the phase structure of QCD, the Nambu--Jona-Lasinio (NJL) model \cite{Nambu:1961_1, Nambu:1961_2, Klevansky:1992, Hatsuda:1994} has been used by many authors.
For example, in Refs.\cite{Buballa:2002, Blaschke:2003, Buballa:2005}, the color superconductivity has been discussed in the NJL model.
Especially, in Ref.\cite{Kitazawa:2002} the authors have included the vector interaction as well as the quark-pairing interaction 
into the NJL model. As a result, it has been shown that the chiral condensate and the color superconducting gap may coexist because of the vector interaction.

The possibility that the spins of quarks may become polarized at large chemical potential has been discussed in Ref.\cite{Tatsumi:2000}.
In addition, since a spin polarization term can be derived from the axial-vector interaction, the authors in Refs.\cite{Tatsumi:2004, Nakano:2003, Maedan:2007} have investigated the possibility of spin polarization by using an axial-vector interaction.
A term similar to the spin polarization term can also be obtained from a tensor interaction in the NJL model, 
which can be interpreted as an anomalous magnetic moment induced dynamically \cite{Ferrer:2014}. 
In this paper that term is called  the ``tensor condensate.''
The tensor condensate from the tensor interaction was investigated in Ref.\cite{Bohr:2012}.
It has been shown in Ref. \cite{Tsue:2012} that the tensor condensed phase may be realized at large chemical potential.
Also, in Refs.\cite{Tsue:2013, Tsue:2015_1}, the relationship between the tensor condensate and color superconductivity has been investigated
at zero temperature.
Moreover, in our preceding papers \cite{Matsuoka:2016, Matsuoka:2017}, the relationship between the chiral condensed phase, ``tensor condensed phase'' and the color superconducting phase 
has been discussed at finite temperature and finite quark chemical potential.
According to these investigations, the chiral condensate, and tensor condensate do not coexist, however, the tensor condensate and two-flavor color superconducting gap 
may coexist at low temperature and large quark chemical potential.
Further, ferromagnetism due to the tensor condensate has been investigated in Ref.\cite{Tsue:2015_2}.
It has been shown that if quarks have an anomalous magnetic moment, the tensor condensate may lead to spontaneous magnetization in high density quark matter.

These conditions, large chemical potential, and low temperature, may be realized in the inner core of compact stars, for e.g., neutron stars and magnetars.
It is known that neutron stars have very strong magnetic fields at their surface \cite{Harding:2006}.
However, nobody understands definitely the mechanism that generates such strong magnetic fields.
We propose that the tensor condensate is the origin of magnetic fields.

Another interesting topic of research related to neutron stars is the difficulty of describing stars with two-solar mass using equation of state (EOS) that includes hyperons or non-nucleonic degrees of freedom \cite{Vidana:2011}.
Though the calculation in the present investigation does not include strange quarks, we will construct compact stars with a quark core using the EOS obtained from the NJL model with the tensor interaction.

In Sec. \ref{lagrangian} we introduce the NJL model with the tensor interaction and then calculate the thermodynamic potential.
In Sec. \ref{numerical} we discuss numerical results and construct hybrid stars.
The last section is devoted to conclusions and remarks.

\section{Lagrangian density and thermodynamic potential}
\label{lagrangian}

In this section we introduce the NJL model with the tensor interaction at finite chemical potential.
In addition to this, we impose the $\beta$ equilibrium and charge neutrality conditions on the system.
The Lagrangian density with flavor $SU(2)$ and color $SU(3)$ symmetry is
\begin{align}
\begin{split}
\mathcal{L}_{\text{total}} &= \mathcal{L}_{\text{NJL}} + \mathcal{L}_T + \mathcal{L}_e + \mathcal{L}_D, \\
\mathcal{L}_{\text{NJL}} &= \bar{\psi} i \gamma^\mu \partial_\mu \psi
+ G_S \big \{ (\bar{\psi} \psi)^2 + (\bar{\psi} i \gamma^5 \vec{\tau} \psi)^2 \big \} \\
\mathcal{L}_T &= -\frac{G_T}{4} \big \{ (\bar{\psi} \gamma^\mu \gamma^\nu \vec{\tau} \psi) \cdot
(\bar{\psi} \gamma_\mu \gamma_\nu \vec{\tau} \psi) \\
& \qquad + (\bar{\psi} i \gamma^5 \gamma^\mu \gamma^\nu \psi)
(\bar{\psi} i \gamma^5 \gamma_\mu \gamma_\nu \psi) \big \}, \\
\mathcal{L}_e &= \bar{\psi}_e i \gamma^\mu \partial_\mu \psi_e, \\
\mathcal{L}_D &= \mu \psi^\dagger \psi 
+ \lambda \bigg \{ \psi^{\dagger}_{e} \psi_e + \frac{1}{3} \psi^{\dagger}_{d} \psi_d - \frac{2}{3} \psi^{\dagger}_{u} \psi_u \bigg \},
\end{split}
\end{align}
where $\tau_i \; (i = 1,2,3)$ is the Pauli matrix that operates in the flavor space and
$
\psi = \begin{pmatrix} \psi_u \\ \psi_d  \end{pmatrix}
$
is the quark fields for up quark $\psi_u$ and down quark $\psi_d$, respectively.
The term $\mathcal{L}_T$ represents the tensor interaction.
\footnote{In other papers, the term $\mathcal{L}_T$ is sometimes written by using $\sigma^{\mu \nu} = \frac{i}{2} [\gamma^\mu, \gamma^\nu]$.
It means that the interaction does not contain terms $\mu = \nu$.
Therefore, in our notation, we do not consider terms $\mu = \nu$.}
We have added the term, $\mathcal{L}_e$, for electrons that neutralize stellar matter.
For simplicity, in this discussion we ignore the current quark mass and the electron mass.
The term $\mathcal{L}_D$ controls densities.
The variables $\mu$ and $\lambda$ handle the quark number density and charge density, respectively.
Namely, $\mu$ is the quark chemical potential, and $\lambda$ will be identified as the electron chemical potential after optimization of the thermodynamic potential.

In this paper, we pay attention to the term, $(\bar{\psi} \gamma^1 \gamma^2  \tau_3 \psi)^2$ in $\mathcal{L}_T$ since using the Dirac representation, we can derive the spin matrix $\Sigma_z = -i \gamma^1 \gamma^2$.
Let us write down the Lagrangian density that we consider in this discussion:
\begin{align}
\begin{split}
\mathcal{L} &= \bar{\psi} i \gamma^\mu \partial_\mu \psi + G_S (\bar{\psi} \psi)^2 + \frac{G_T}{2} (\bar{\psi} \Sigma_z \tau_3 \psi)^2 \\
&+ \bar{\psi}_e i \gamma^\mu \partial_\mu \psi_e + \mu \psi^\dagger \psi \\
&+ \lambda \bigg ( \psi^\dagger_e \psi_e + \frac{1}{3} \psi^\dagger_d \psi_d - \frac{2}{3} \psi^\dagger_u \psi_u \bigg ).
\end{split}
\end{align}

In the mean field approximation, the above Lagrangian density becomes
\begin{align}
\begin{split}
\mathcal{L}_{\text{MFA}} = & \sum_{f = u,d} \bar{\psi}_f \Big[ 
i \gamma^\mu \partial_\mu - M - \hat{f} F \Sigma_z + \mu_f \gamma^0 \Big] \psi_f \\
& -\frac{M^2}{4 G_S} -\frac{F^2}{2 G_T} 
+ \bar{\psi}_e \Big [i \gamma^\mu \partial_\mu + \lambda \gamma^0 \Big ] \psi_e,
\end{split}
\end{align}
where 
\begin{gather}
M = -2 G_S (\langle \bar{\psi}_u \psi_u \rangle + \langle \bar{\psi}_d \psi_d \rangle), \\
F = -G_T (\langle \bar{\psi}_u \Sigma_z \psi_u \rangle - \langle \bar{\psi}_d \Sigma_z \psi_d \rangle),
\end{gather}
and $\langle \cdot \rangle$ means expectation value.
Note that the variables, $M$ and $F$, are the order parameters, the chiral condensate, and tensor condensate, respectively.
In order to simplify the notation, we have defined $\hat{f}$ in the following way:
\begin{align*}
\hat{f} = \begin{cases} 1 & \text{for} \; f = u \\ -1 & \text{for} \; f = d \end{cases},
\end{align*}
where $u$ and $d$ represent up quark and down quark, respectively.
In addition, we also introduce
\begin{align*}
\mu_u = \mu - \frac{2}{3} \lambda, \quad
\mu_d = \mu + \frac{1}{3} \lambda.
\end{align*}
We will realize the above variables as chemical potential of the up quark and down quark, respectively.
By using the standard technique, we can obtain the single-particle energy as follows:
\begin{align}
\begin{split}
\epsilon^\alpha &= \sqrt{p_z^2 + \Big ( \sqrt{p_x^2+p_y^2+M^2} + \alpha F \Big ) ^2}, \\
\epsilon_e &= p,
\end{split}
\end{align}
where $\alpha = \pm 1$.
The first one is for quarks and the second one is for electrons.
As we have commented, $\lambda$ is identified as the electron chemical potential.
Therefore, we write $\mu_e := \lambda$ in the following discussion.
The thermodynamic potential $\Phi$ is obtained as
\begin{widetext}
\begin{align}
\begin{split}
&\Phi(M,F,\mu,\mu_e) = \Phi_q + \Phi_e, \\
&\Phi_q = N_C \int_{|\vec{p}| \le \Lambda} \frac{d^3 p}{(2 \pi)^3}
\sum_{f = u,d \atop \alpha = \pm 1} \bigg \{ 
\Big ( \epsilon^\alpha - \mu_f \Big) \theta(\mu_f - \epsilon^\alpha) - \epsilon^\alpha \bigg \}
+ \frac{M^2}{4 G_S} + \frac{F^2}{2 G_T}, \\
&\Phi_e = 2\int_{-\infty}^\infty \frac{d^3 p}{(2 \pi)^3} \Big ( \epsilon_e - \mu_e \Big ) \theta(\mu_e - \epsilon_e) = - \frac{\mu_e^4}{12 \pi^2},
\end{split}
\end{align}
\end{widetext}
where $\Phi_q$ and $\Phi_e$ are the thermodynamic potentials for the quark and electron, respectively, and $N_C = 3$ is the number of color.
We have introduced a three-momentum cutoff parameter $\Lambda$ in the quark thermodynamic potential.

Here we discuss renormalization of the thermodynamic potential briefly.
We can derive the pressure $P$ as $P = -\Phi$.
However, in this case, 
\begin{align*}
P(\mu = \mu_e = 0) = -\Phi(M_0,F_0,0,0) \neq 0.
\end{align*}
where $M_0$ and $F_0$ minimize $\Phi$ when $\mu = \mu_e = 0$.
Namely, the pressure is not zero at zero chemical potentials.
Therefore, we redefine a new thermodynamic potential as
\begin{align}
\begin{split}
&\Phi_R (M,F,\mu,\mu_e) \\
&:= \Phi(M,F,\mu,\mu_e) - \Phi(M_0,F_0,0,0).
\end{split}
\end{align}
Then, the renormalized pressure is computed as $P = -\Phi_R$.

We can calculate the quark number density by differentiating the thermodynamic potential with respect to the quark chemical potential:
\begin{align}
\rho = - \frac{\partial \Phi_R}{\partial \mu} = \rho_u + \rho_d,
\end{align}
where $\rho_u$ and $\rho_d$ are the number density of up quark and down quark, respectively.
On the other hand, we can obtain the electron number density in the following way:
\begin{align}
\rho_e = -\frac{\partial \Phi_e}{\partial \mu_e} = \frac{\mu_e^3}{3 \pi^2}.
\end{align}

We minimize the thermodynamic potential under the condition $\partial \Phi_R / \partial \mu_e = \partial \Phi_R / \partial \lambda = 0$, namely,
\begin{align}
\rho_e + \frac{1}{3} \rho_d - \frac{2}{3} \rho_u = 0.
\end{align}
It means the charge neutrality condition.
Using the above condition and the definition of the quark number density, the number density of up quark and down quark can be written as
\begin{align}
\rho_u = \frac{1}{3} \rho + \rho_e, \quad
\rho_d = \frac{2}{3} \rho - \rho_e.
\end{align}
In addition, the chemical potential of the up quark and down quark can be written as
\begin{align}
\mu_u = \mu - \frac{2}{3} \mu_e, \quad
\mu_d = \mu + \frac{1}{3} \mu_e.
\end{align}
Then we can derive $\mu_d = \mu_u + \mu_e$, and it is the $\beta$ equilibrium condition.

We can write down energy density through the thermodynamic relation as
\begin{equation}
E = \Phi_R + \mu \rho = \Phi_R + \mu (\rho_u + \rho_d).
\end{equation}

\section{Numerical results}
\label{numerical}

We comment on the parameters.
The NJL model is not a renormalizable theory, therefore, we have introduced a three-momentum cutoff parameter, $\Lambda$.
The parameters, $\Lambda$ and $G_S$, are determined to reproduce the dynamical quark mass and the pion decay constant in the vacuum.
The value of the coupling constant with tensor interaction $G_T$ may be determined by experimental values \cite{Jaminon:1998, Jaminon:2002} 
or the Fierz transformation of the scalar and pseudoscalar interaction in the NJL model.
In these cases, the sign of $G_T$ becomes opposite to our model.
On the other hand, in Ref. \cite{Battistel:2016}, the authors use the opposite and same sign.
The value of $G_T$ is not well known, therefore, we treat $G_T$ as a free parameter.
The values of parameters are enumerated in the Table. I.

In the following discussion, we do not use the quark chemical potential but baryon chemical potential.
It is defined in the following way:
\begin{align}
\mu_B = 3 \mu.
\end{align}
In addition, we also introduce the baryon number density as
\begin{align}
\rho_B = \frac{1}{3} \rho.
\end{align}

\begin{table}
\centering
\caption{Parameter set}

\begin{tabular}{cccc}
\hline
Model & $\Lambda \; [\text{GeV}]$ & $G_S \; [\text{GeV}^{-2}]$ & $G_T \; [\text{GeV}^{-2}]$ \\ \hline \hline
GT0 & $0.631$ & $5.5$ & $0$ \\ \hline
GT11 & $0.631$ & $5.5$ & $11.0$ \\ \hline
GT12 & $0.631$ & $5.5$ & $12.0$ \\ \hline
GT13 & $0.631$ & $5.5$ & $13.0$ \\ \hline
GT14 & $0.631$ & $5.5$ & $14.0$ \\ \hline
\end{tabular}

\end{table}

\subsection{Behavior of $F$}

\begin{figure}
\centering
\includegraphics[width=\columnwidth]{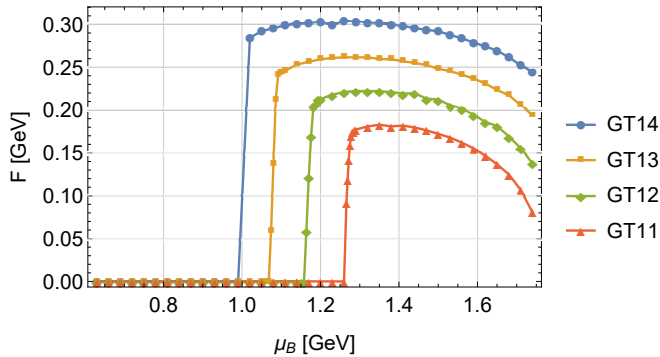}
\caption{The relationship between the baryon chemical potential $\mu_B$ and the tensor condensate $F$ is shown.
The horizontal and vertical axes represent the baryon chemical potential and tensor condensate, respectively.}
\label{CPvsF}
\end{figure}

\begin{figure*}
\centering
\includegraphics[width = 1.5 \columnwidth]{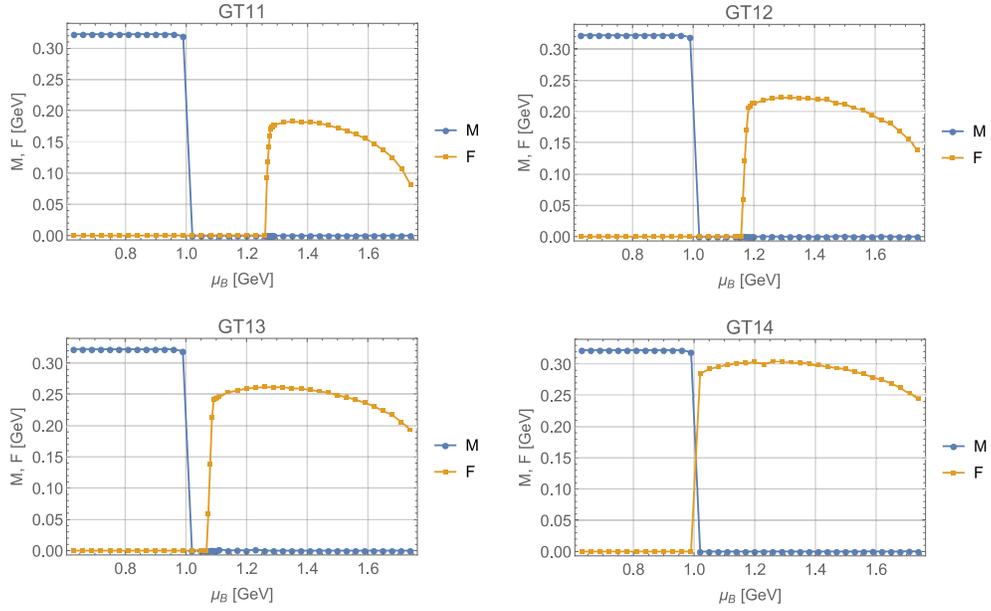}
\caption{These four figures show the competition between the chiral condensate $M$ and the tensor condensate $F$ in model GT11, GT12, GT13, and GT14, respectively.
The horizontal axis is the baryon chemical potential $\mu_B$ and the vertical axis is the values of $M$ and $F$.}
\label{competition}
\end{figure*}

Figure \ref{CPvsF} shows the behavior of the tensor condensate $F$.
We vary the value of baryon chemical potential from 0.6 GeV to 1.8 GeV.
Since the model GT0 does not have the tensor interaction, the tensor condensate does not occur.
Thus, the model GT0 does not appear in this figure.
The figure shows that as the $G_T$ becomes larger, $F$ can get nonzero values at smaller chemical potential.
Moreover, the value of $F$ becomes larger.

Next we show the competition between the chiral condensate and the tensor condensate;
see Fig. \ref{competition}.
The four graphs represent the competition in models GT11, GT12, GT13, and GT14, respectively.
The horizontal axis is the baryon chemical potential, and the vertical axis is the chiral condensate $M$ and the tensor condensate $F$.
In all models, while $\mu_B$ is small enough, the chiral condensate is realized.
Then, when $\mu_B$ becomes large enough, the tensor condensate is realized.
In our calculation, we cannot obtain the situation $M \neq 0$ and $F \neq 0$.
Let us summarize the phase transition in our model here.
When we use models GT11, GT12, and GT13, the phase transition is of the type 
\begin{align*}
&\text{Chiral condensed phase} \; (M \neq 0, \; F = 0) \\
& \quad \longrightarrow \text{Chiral symmetric phase} \; (M = 0, \; F = 0) \\
& \qquad \longrightarrow \text{Tensor condensed phase} \; (M = 0, \; F \neq 0),
\end{align*}
as the baryon chemical potential becomes larger.
The phase transition occurs via the chiral symmetric phase.
On the other hand, when we use model GT14, the phase transition is of the type
\begin{align*}
&\text{Chiral condensed phase} \; (M \neq 0, \; F = 0)\\
& \quad \longrightarrow \text{Tensor condensed phase} \; (M = 0, \; F \neq 0),
\end{align*}
as the baryon chemical potential becomes larger.

\subsection{Behaviors of $\rho$, $\mu_e$ and etc}

\begin{figure}
\centering
\includegraphics[width=\columnwidth]{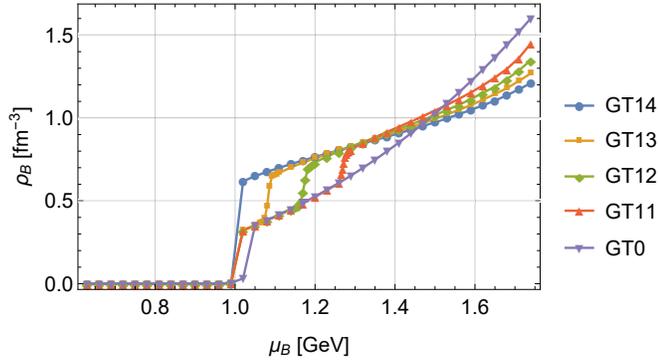}
\caption{The relationship between the baryon chemical potential $\mu_B$ and the baryon number density $\rho_B$ is shown.
The horizontal and vertical axes represent the baryon chemical potential and baryon number density, respectively.}
\label{CPvsD}
\end{figure}

\begin{figure}
\centering
\includegraphics[width = \columnwidth]{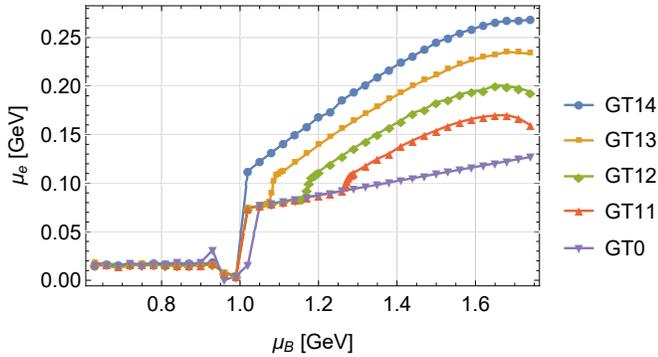}
\caption{This figure shows the relation between the baryon chemical potential $\mu_B$ and the electron chemical potential $\mu_e$.
The horizontal axis is $\mu_B$ and the vertical axis is $\mu_e$.}
\label{CPvsCPe}
\end{figure}

\begin{figure}
\centering
\includegraphics[width=\columnwidth]{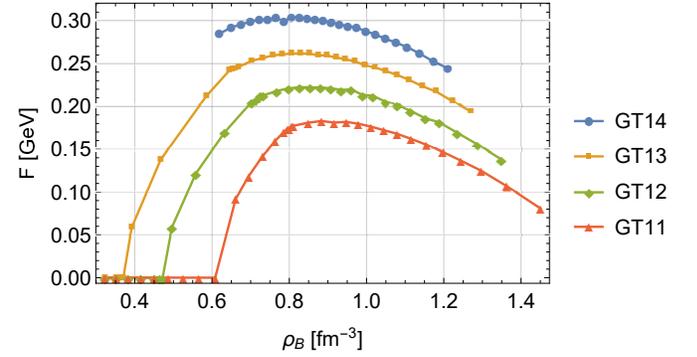}
\caption{The relationship between the baryon number density $\rho_B$ and the tensor condensate $F$ is shown.
The horizontal and vertical axes represent the baryon number density and the tensor condensate, respectively.}
\label{DvsF}
\end{figure}

\begin{figure}
\centering
\includegraphics[width = \columnwidth]{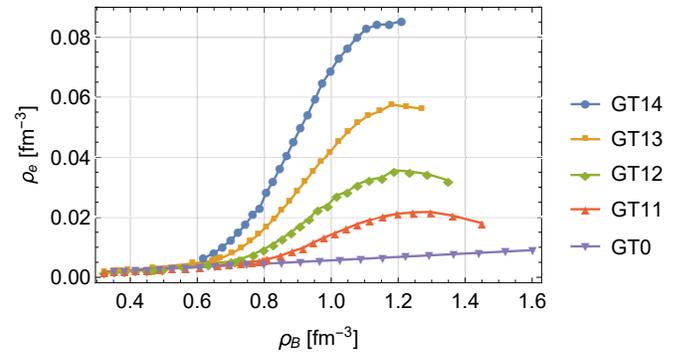}
\caption{This figure shows the relation between the baryon number density $\rho_B$ and the electron number density $\rho_e$.
The horizontal and vertical axes represent $\rho_B$ and $\rho_e$, respectively.}
\label{DvsDe}
\end{figure}

\begin{figure}
\centering
\includegraphics[width = \columnwidth]{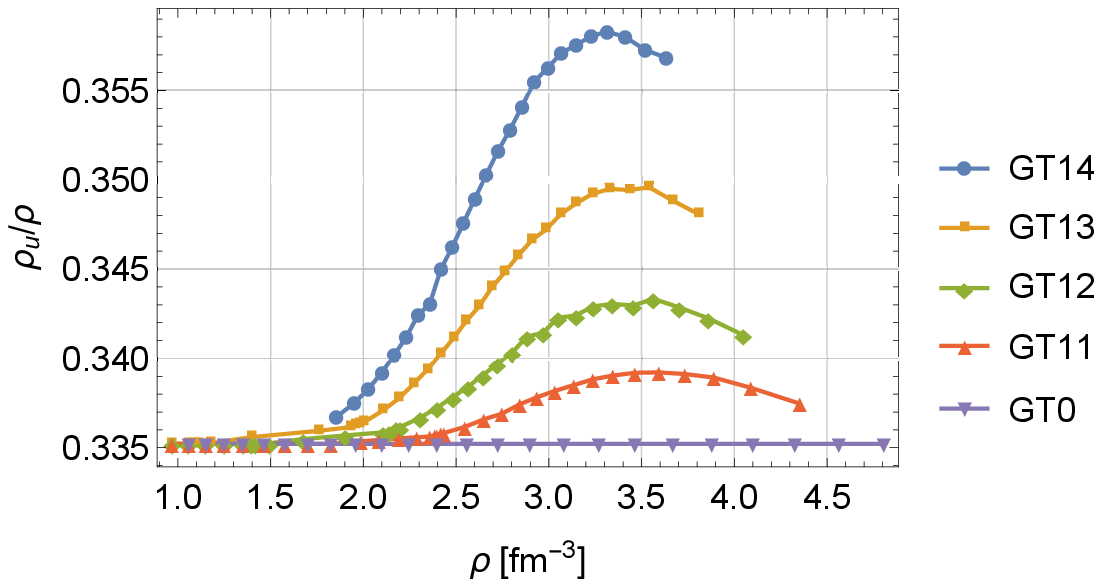}
\caption{This figure shows the proportion of up quarks in the system.
The horizontal axis is the quark number density $\rho$ not the baryon number density.
The vertical axis is $\rho_u / \rho$.}
\label{fraction_1}
\end{figure}

\begin{figure}
\centering
\includegraphics[width = \columnwidth]{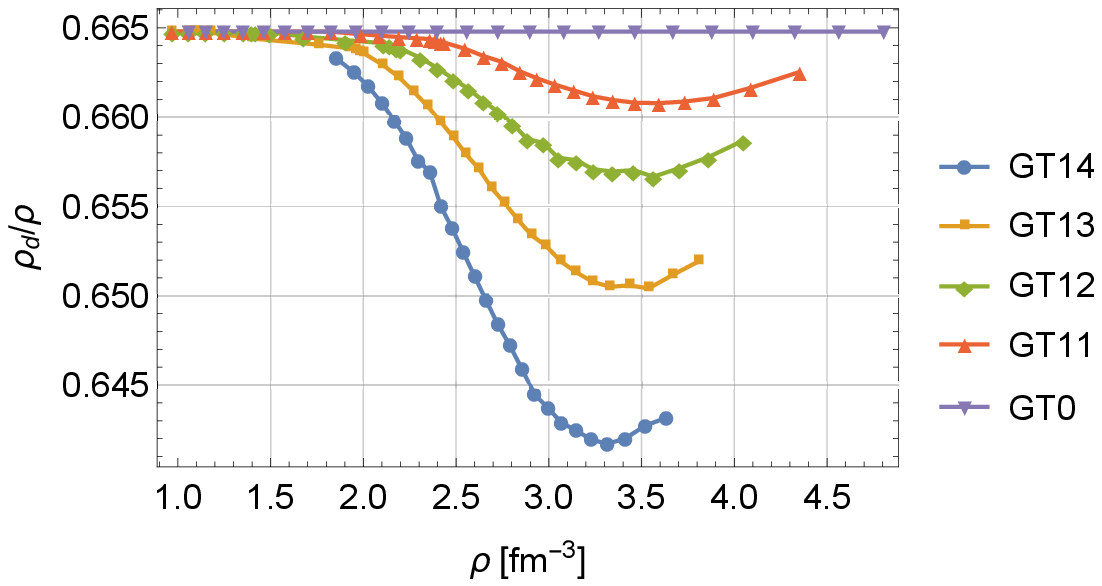}
\caption{This figure shows the proportion of down quarks in the system.
The horizontal axis is the quark number density $\rho$ not the baryon number density.
The vertical axis is $\rho_d / \rho$.}
\label{fraction_2}
\end{figure}

\begin{figure}
\centering
\includegraphics[width=\columnwidth]{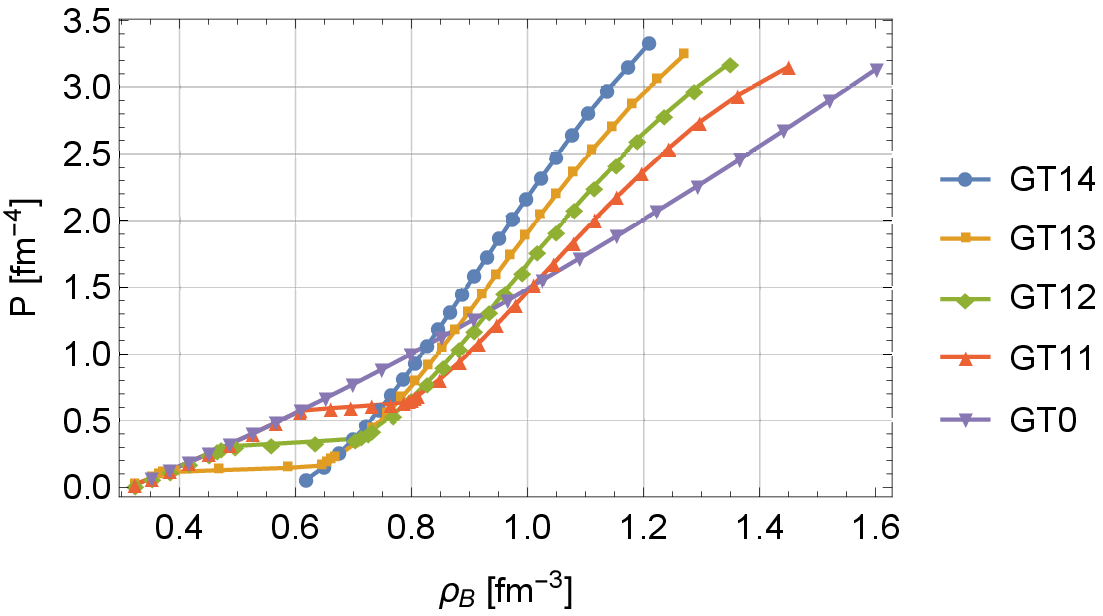}
\caption{The relationship between the baryon number density $\rho_B$ and pressure $P$ is shown.
The horizontal and vertical axes represent the baryon number density and pressure, respectively.}
\label{DvsP}
\end{figure}

Figure \ref{CPvsD} shows the behavior of the baryon number density of the system.
The horizontal and vertical axes are the baryon chemical potential and the baryon number density, respectively.
Model GT0 has one discontinuity at $\mu_B \sim 1.0 \; \text{GeV}$.
It means that the phase transition occurs from the chiral condensed phase to the chiral symmetric phase.
The models GT11, GT12, and GT13 have one discontinuity at $\mu_B \sim 1.0 \; \text{GeV}$ and one sharp rise, but continuous, in the value.
The former discontinuity corresponds to the phase transition from the chiral condensed phase to the chiral symmetric phase.
The latter sharp rise corresponds to the phase transition from the chiral symmetric phase to the tensor condensed phase.
On the other hand, model GT14 has only one discontinuity.
The discontinuity corresponds to the phase transition from the chiral condensed phase to the tensor condensed phase.

Next let us see the behavior of the electron chemical potential; see Fig. \ref{CPvsCPe}.
The horizontal and vertical axes represent the baryon chemical potential and the electron chemical potential, respectively.
In all models, the values of $\mu_e$ are nonzero at $\mu_B = 0$ and small baryon chemical potentials.
We consider that this is a numerical error, the true values are zero while $\mu_B \lesssim 1.0 \; \text{GeV}$.
Although $\mu_e \neq 0$, this occurs at very low baryonic densities, out of the range of the baryonic densities at the transition: hadronic matter $\leftrightarrow$ quark matter, and below the densities needed to build the hybrid star's EOS.
As we have seen in Fig. \ref{CPvsD}, each of models GT0 and GT14 has one discontinuity, and each of models GT11, GT12, and GT13 has one discontinuity and one continuous sharp rise in the values of $\mu_e$.
Here we comment on the discontinuities.
It looks like the discontinuities in models GT11, GT12, and GT13 are at smaller baryon chemical potential than that in model GT0.
However, the discontinuities in model GT0, ..., and GT13 should be coincident since the tensor condensate has not appeared yet.
The differences are due to numerical problem, which, however, will not affect the main conclusions on the hybrid star structure.

Figure \ref{DvsF} shows the relationship between the baryon number density and the tensor condensate.
The horizontal and vertical axes are $\rho_B$ and $F$, respectively.
For the models GT11, GT12, and GT13, $F$ does not have a finite value at small baryon number densities.
However, if the baryon number density becomes large enough, $F$ can get nonzero values.
Model GT14 does not have points at small baryon number density because $\rho_B$ with such values does not occur (see Fig. \ref{CPvsD}).
Of course, $F$ is zero at $\rho_B = 0$.
If the baryon number density exceeds $\rho_B \sim 0.6 \; \text{fm}^{-3}$, $F$ can obtain finite values.

In Fig. \ref{DvsDe}, the relation between the baryon number density and the electron number density is shown.
The horizontal axis is $\rho_B$, and the vertical axis is $\rho_e$.
In model GT0, the value of $\rho_e$ increases linearly as $\rho_B$ becomes larger.
On the other hand, in models GT11, ..., and GT14, the curves of $\rho_e$ rise sharply after the tensor condensate is realized.
In addition, the value of $\rho_e$ becomes larger at fixed $\rho_B$ as the value of $G_T$ becomes larger.
This is due to the fact that the tensor condensate favors larger (smaller) up-quark (down-quark) fractions, see discussion below, and in order to ensure electric charge neutrality the electron number density must increase.

In Figs. \ref{fraction_1} and \ref{fraction_2}, we plot up-quark and down-quark fractions, $\rho_u / \rho$ and $\rho_d / \rho$, as a function of the quark number density $\rho$.
In model GT0, the two fractions do not change very much.
In other models, because of the tensor condensate, $\rho_u / \rho$ increases and $\rho_d / \rho$ decreases.
The variations of the two fractions become larger as $G_T$ becomes larger.
The tensor interaction reduces the energy of the system and it energetically favors a reduction of the down-quark Fermi momenta.
The energy gain compensates the increase of the electric density.

In Fig. \ref{DvsP}, the baryon number density versus pressure plot is depicted.
We have eliminated some isolated points at $P = \rho_B = 0$ for numerical reasons when we compute interpolated functions.
Therefore, the baryon number density $\rho_B$ starts from a finite value instead of zero.
The pressure of the model GT0 increases monotonically from the origin.
\footnote{
It should be noted that the origin in this graph is not $(\rho_B,P) = (0,0)$.}
On the other hand, the models GT11, GT12, and GT13 have plateaus.
These plateaus indicate the onset of the tensor condensate.

\subsection{Hybrid star}

First, we explain what we refer to as a ``hybrid star.''
The hybrid star has an inner core that consists of the quark matter, and the outer core and crust consist of hadrons.
In order to obtain the EOS of hadrons, we use the NL3$\omega \rho$ model \cite{Pereira:2016}.

\begin{figure}
\centering
\includegraphics[width=\columnwidth]{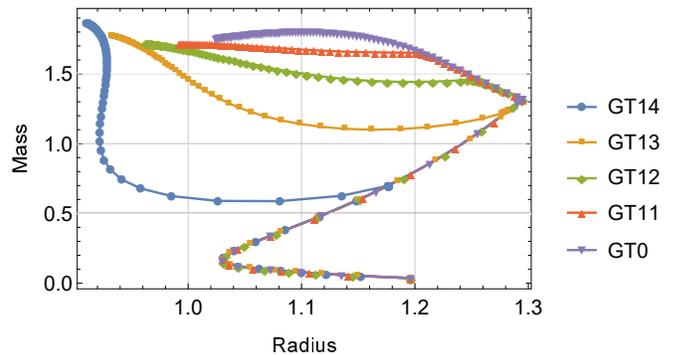}
\caption{The relationship between the radius and mass of hybrid stars is shown.
We have normalized the mass $m$ by the solar mass $m_\text{sun}$ and the radius $r$ by $r_0 = 10 \; \text{km}$.
The horizontal and vertical axes represent the radius $r / r_0$ and mass $m / m_{\text{sun}}$, respectively.}
\label{RvsM}
\end{figure}

Let us discuss the relationship between the radius and mass of hybrid stars by solving the Tolman-Oppenheimer-Volkoff (TOV) equation numerically.
In Fig. \ref{RvsM}, the radius ($r$)-mass ($m$) relation is depicted.
We have normalized $r$ and $m$ by $r_0 = 10 \, \text{km}$ and the solar mass $m_\text{sun}$, respectively.
See the models GT0, GT11, and GT12.
The curves bend at $(r/r_0, m/m_{\text{sun}}) \sim (1.3, 1.3)$.
The point corresponds to the appearance of the quark matter at the inner core of the hybrid star; the curves under this point have ``hadron cores,'' on the other hand, the curves above this point have ``quark cores.''
The curves of the models GT11 and GT12 bend at $(r/r_0, m/m_{\text{sun}}) \sim (1.2, 1.7)$ and $(r/r_0, m/m_{\text{sun}}) \sim (1.25, 1.5)$ again, respectively.
This point means that the tensor condensate appears at the core of the hybrid stars; the curves under this point have cores which are $M = F = 0$.
The curves above this point have cores which are $M = 0$ and $F \neq 0$.
The curve of the model GT13 snaps off at $(r/r_0, m/m_{\text{sun}}) \sim (1.28, 1.2)$.
This point means the onset of the tensor condensate at the core.
Since the tensor condensate is realized at smaller chemical potential as $G_T$ becomes larger, the curve of the model GT14 bends earlier than the model GT13.
The ``bending point'' for the model GT14 is at $(r/r_0, m/m_{\text{sun}}) \sim (1.18, 0.7)$.
We note that models GT13 and GT14 predict ``twin stars'' \cite{Benic:2015, Alvarez:2017}, i.e., stable stars with the same mass but very different radii: a hadronic star and a hybrid star, the first one with a radius is 2 km larger than the hybrid star radius.
For more discussion concerning twin stars, please see Refs. \cite{Benic:2015, Alvarez:2017}.

\subsection{Estimation of magnetic moment}

We estimate the magnetic moment density by using Eq. (22) in Ref. \cite{Maruyama:2018}.
In our case, the expression of the magnetic moment density becomes
\begin{align}
M_{\text{mag}} = \bigg ( \frac{2}{3} \bar{\rho}_u + \frac{1}{3} \bar{\rho}_d \bigg ) \frac{e}{2 m_q} \frac{\langle \bar{\psi} i \gamma^1 \gamma^2 \tau_3 \psi \rangle}{\langle \psi^\dagger \psi \rangle} 3 \rho_B,
\end{align}
where $\bar{\rho}_u$ and $\bar{\rho}_d$ are the fraction of up and down quark, respectively.
We use the current quark mass: $m_q = 0.005 \; \text{GeV}$.
In the above equation, $\bar{\rho}_u$ and $\bar{\rho}_d$ satisfy $\bar{\rho}_u + \bar{\rho}_d = 1$; thus, we can write
\begin{align*}
\bar{\rho}_u = \frac{\rho_u}{\rho_u + \rho_d}, \;
\bar{\rho}_d = \frac{\rho_d}{\rho_u + \rho_d}.
\end{align*}
In Ref. \cite{Maruyama:2018}, a relation $\bar{\rho}_u = \bar{\rho}_d$ is used.
However, in our case, this condition is not satisfied because we are imposing the $\beta$ equilibrium and the charge neutrality conditions on the system.
Thus, we must use numerical data for $\bar{\rho}_u$ and $\bar{\rho}_d$, namely, $\rho_u$ and $\rho_d$.
By using the following relations: $3 \rho_B = \rho$, $\rho_u = \rho / 3 + \rho_e$ and $\rho_d = 2 \rho / 3 - \rho_e$, we can transform $M_{\text{mag}}$ into
\begin{align*}
M_{\text{mag}} = (4 \rho_B + \rho_e) \times \frac{e F}{18 m_q G_T \rho_B}.
\end{align*}
Here we discuss the dimension.
We are using the following unit: $c = \hbar = \mu_0 = 1$, where $\mu_0$ is the vacuum permeability.
Therefore, the dimension of $M_{\text{mag}}$ is $[\text{GeV}^{2}]$.
We also have the relation: $1 \; \text{Gauss} = 1.955 \times 10^{-20} \; \text{GeV}^2$.
Thus, the dimension of $M_{\text{mag}}$ is [Gauss].

\begin{figure}
\centering
\includegraphics[width=\columnwidth]{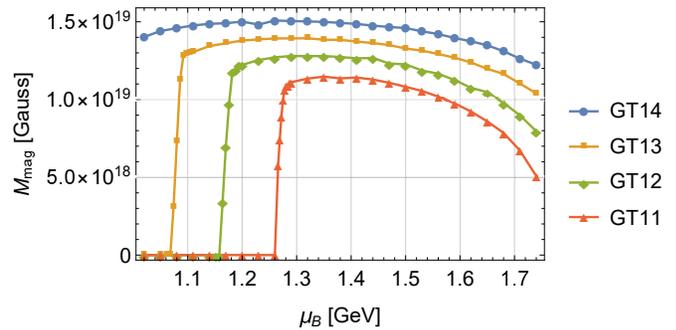}
\caption{The figure shows the relationship between the chemical potential and the magnetic moment.
The vertical and horizontal axes represent the magnetic moment density and the baryon chemical potential, respectively.}
\label{CPvsMmag}
\end{figure}

See Fig. \ref{CPvsMmag}.
The vertical and horizontal axes are the magnetic moment density and the baryon chemical potential, respectively.
As $G_T$ becomes larger, $M_{\text{mag}}$ gets larger and becomes finite at smaller baryon chemical potentials.
We can get $M_{\text{mag}} \sim 10^{19} \; \text{G}$.

\section{Conclusions and remarks}
\label{conclusions}

In this paper we have investigated the behavior of tensor condensate and its implication on the properties of the hybrid star by using the NJL model with the tensor interaction under the $\beta$ equilibrium and charge neutrality conditions.
As the value of $G_T$ becomes larger, the value of $F$ increases.
Moreover, $F$ has nonzero values at a smaller baryon chemical potential.

In addition, we have constructed hybrid stars by using the EOS of this model.
When $G_T \le 14.0 \; \text{GeV}^{-2}$, we obtain no hybrid stars with two-solar mass.
However, if we include the vector interaction as repulsion, we may obtain compact stars with two-solar mass.
It is outside the scope of this paper.
We have built stars with a polarized core, and a finite tensor condensate.
This could be a mechanism that explains the strong magnetic fields inside magnetars.
We expect that our scenario is valid also for hybrid stars with $m_\text{hybrid} > 2 m_\text{sun}$, where $m_\text{hybrid}$ is the mass of the hybrid star.

We must, however, point out some problems that have arisen while calculating the hybrid star families.
See the radius-mass curves in Fig. \ref{RvsM}.
The curves do not reach the maximum mass.
In order to depict the curves, we have used numerical EOS data with baryon chemical potential, $\mu_B \lesssim 1.8 \; \text{GeV}$, namely, $\mu \lesssim 0.6 \; \text{GeV}$.
Since the quark Fermi momentum should not exceed the value of cutoff parameter $\Lambda$ it is difficult to extend the curves any more.
If we use a larger cutoff parameter, we may extend the radius-mass curve, but this requires the determination of a new set of parameters with a large cutoff and this will not be considered in this present work.

In this research, we use the flavor $SU(2)$ NJL model with the tensor interaction.
Many researchers consider that the color superconducting phase may be realized in the high density region.
In addition, at large chemical potential we should not ignore the contribution from strange quarks.
Thus, it is interesting to extend our model to flavor $SU(3)$ case.
In our discussion, the renormalization was defined in a such a way that the pressure vanishes when the quark and electron chemical potential are zero.
However, a different renormalization procedure of the pressure could have been carried out as done in \cite{Pereira:2016}.
These will be considered in a future work.

\section*{ACKNOWLEDGMENTS}

H.M. and Y.T. would like to express their sincere thanks to 
the members of Many-Body Theory Group of Kochi University.
This work  was  partially supported  by  “Fundacao para  a Ciencia  e  Tecnologia”,  Portugal,  under  the  projects No. UID/FIS/04564/2016,  POCI-01-0145-FEDER-029912  [with  financial support  from  POCI,  in  its FEDER  component,  and  by  the FCT/MCTES  budget through  national  funds  (OE)].

\appendix

\section{How to construct hybrid stars}

In this Appendix, we discuss how to construct hybrid stars.
As we have referred, the hybrid star contains quarks in the inner core while the outer core consists of hadrons.
In order to obtain the EOS of hadrons, we use the NL3$\omega \rho$ model.
The Lagrangian density is obtained as
\begin{widetext}
\begin{align*}
\mathcal{L}_{\text{NL3$\omega \rho$}} &= 
\sum_{N=p,n} \bar{\psi}_N \bigg [ 
\gamma^\mu (i \partial_\mu - g_{\omega N} \omega_\mu  - \frac{1}{2} g_{\rho N} \bm{\tau} \cdot \bm{\rho}_\mu) - (m_N - g_{\sigma N} \sigma)
\bigg ] \psi_N \\
& + \frac{1}{2} \partial_\mu \sigma \partial^\mu \sigma - \frac{1}{2} m^{2}_{\sigma} \sigma^2 
-\frac{1}{4} \Omega_{\mu \nu} \Omega^{\mu \nu} + \frac{1}{2} m^{2}_{\omega} \omega^\mu \omega_\mu
-\frac{1}{4} \bm{\rho}^{\mu \nu} \cdot \bm{\rho}_{\mu \nu} + \frac{1}{2} m^{2}_{\rho} \bm{\rho}^\mu \cdot \bm{\rho}_\mu \\
& -\frac{1}{3} b m_N (g_{\sigma N} \sigma)^3 -\frac{1}{4} c (g_{\sigma N} \sigma)^4
+ \Lambda_\omega (g^{2}_{\omega} \omega_\mu \omega^\mu)(g^{2}_{\rho} \bm{\rho}_\mu \cdot \bm{\rho}^\mu),
\end{align*}
\end{widetext}
where $\Omega_{\mu \nu} = \partial_\mu \omega_\nu - \partial_\nu \omega_\mu$ and $\bm{\rho}_{\mu \nu} = \partial_\mu \bm{\rho}_\nu - \partial_\nu \bm{\rho}_\mu$.
The Lagrangian density contains the fields of nucleons ($\psi_p$ and $\psi_n$), $\sigma$ meson ($\sigma$), $\rho$ meson ($\bm{\rho}^\mu$) and $\omega$ meson ($\omega^\mu$).
We note that the EOS obtained by this model can hold compact stars with two-solar mass.

Our strategy for numerical calculation is as follows:
\begin{enumerate}
\item Give an arbitrary value to the central energy density of the hybrid star.
\item If the value is large enough that the quark matter is realized, go to step 3.
If the value is not enough, jump to step 5.
\item Solve the TOV equation from the center to the outside of the star by using the EOS of quarks until a certain reference pressure obtained by performing a Maxwell construction.
\item At the reference pressure, switch the EOS to the one of hadrons.
\item Solve the TOV equation to the outside of the star with the EOS of hadrons until the pressure of the star vanishes.
\item Change the value of central energy density and go back to step 2.
\end{enumerate}

\begin{figure*}
\centering
\includegraphics[width=1.8 \columnwidth]{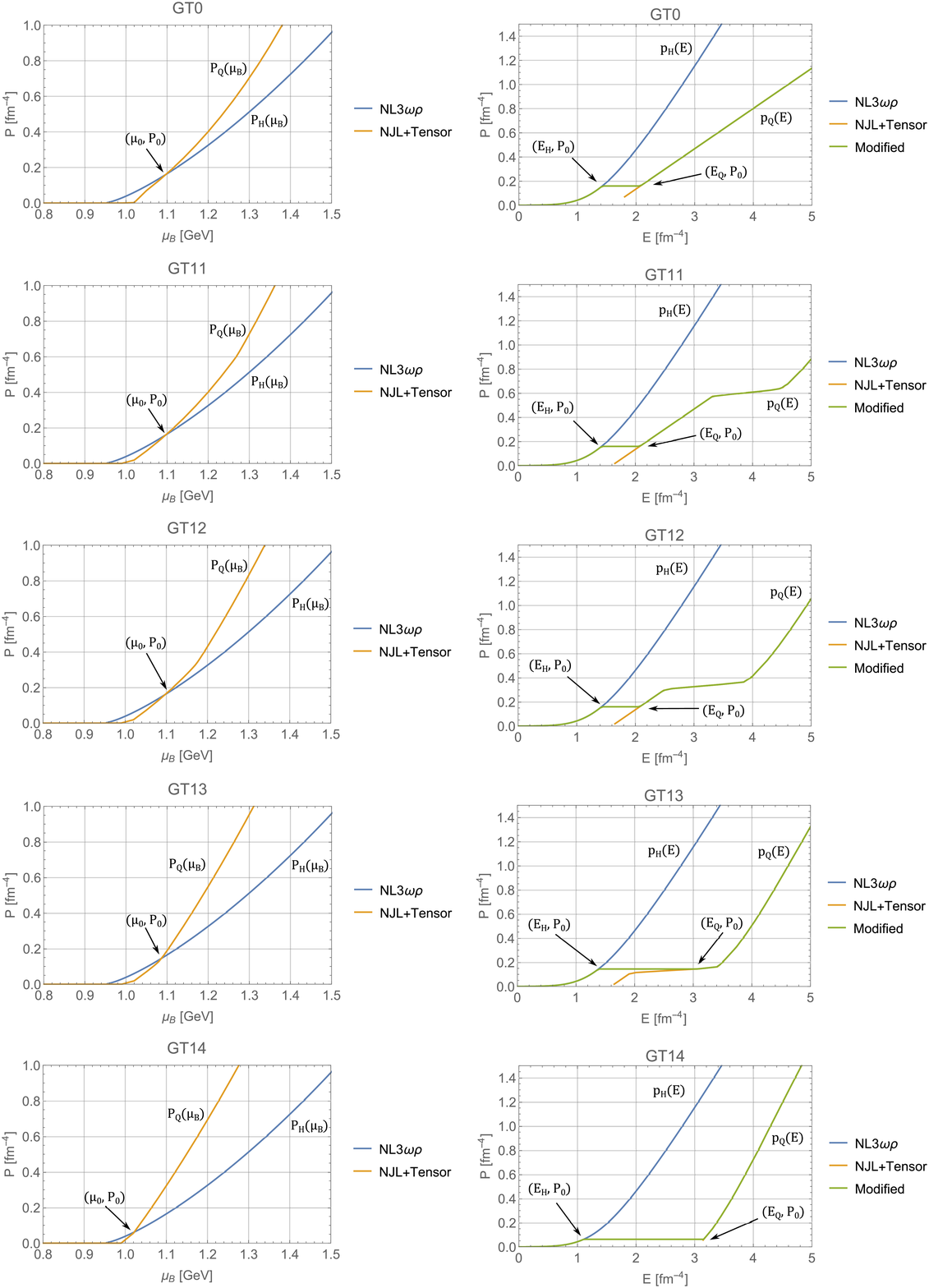}
\caption{
In left column, the horizontal and vertical axes represent the baryon chemical potential and pressure, respectively.
In right column, the horizontal and vertical axes are the energy density and pressure, respectively.
}
\label{EOS}
\end{figure*}

In addition to the above strategy, we should discuss the way to switch the EOS from quarks to hadrons.
In Fig. \ref{EOS}, we plot the EOS for the models GT0, GT11, ..., and GT14.
The figures in the left column represent baryon chemical potential-pressure plots, and those in the right column represent energy density-pressure plots.
The blue curves (NL3$\omega \rho$) are the EOS obtained by the NL3$\omega \rho$ model.
The yellow curves (NJL+Tensor) represent the EOS obtained by the NJL model with the tensor interaction.
In addition, the green curves (Modified) are the EOS that is used in the construction of hybrid stars.
We introduce some functions: $P_i(\mu), \; p_i(E)$ and $(i=Q,H)$.
The functions $P_{Q}$ ($P_H$) and $p_{Q}$ ($p_H$) refer to the pressure of NJL+Tensor (NL3$\omega \rho$).
The argument of $P_i \; (i=Q,H)$ is the baryon chemical potential.
On the other hand, the argument of $p_i \; (i=Q,H)$ is the energy density.

In the left column, there are the cross points, where $P_Q(\mu_0) = P_H(\mu_0) = P_0$.
We switch from EOS to the other at the pressure $P_0$.
This is the Maxwell construction.
In the right column, the horizontal part of the green curve, $E_H \le E \le E_Q$, is defined so that $p_H(E_H) = p_Q(E_Q) = P_0$.

\clearpage




\end{document}